\documentclass[conference]{IEEEtran}
\IEEEoverridecommandlockouts
\usepackage{cite}
\usepackage{amsmath,amssymb,amsfonts}
\usepackage{algorithmic}
\usepackage{graphicx}
\usepackage{textcomp}
\usepackage{xcolor}
\usepackage[caption=false,font=footnotesize]{subfig} 
\usepackage{acro}
\usepackage{balance}
\makeatletter
\renewcommand{\fnum@figure}{\textbf{\figurename~\thefigure}}
\makeatother
\usepackage[caption=false,font=footnotesize]{subfig}

\DeclareAcronym{ue}{
  short = UE,
  long  = user equipment,
}
\DeclareAcronym{mimo}{
  short = MIMO,
  long  = multiple-input multiple-output,
}
\DeclareAcronym{thz}{
  short = THz,
  long  = Terahertz,
}
\DeclareAcronym{ap}{
  short = AP,
  long  = access point,
}
\DeclareAcronym{upa}{
  short = UPA,
  long  = uniform planar array,
}
\DeclareAcronym{snr}{
  short = SNR,
  long  = signal-to-noise ratio,
}
\DeclareAcronym{rmse}{
  short = RMSE,
  long  = root-mean-squared error,
}
\DeclareAcronym{se}{
  short = SE,
  long  = spectral efficiency,
}

\begin{document}

\title{Bessel Beam Optimization for Near-Field THz Communications under UE Location Uncertainty\vspace{-3mm}}

\author{\IEEEauthorblockN{Aditya Jolly\IEEEauthorrefmark{1}\IEEEauthorrefmark{2},
Vitaly Petrov\IEEEauthorrefmark{2}, Gábor Fodor\IEEEauthorrefmark{1}\IEEEauthorrefmark{2} and
Emil Björnson\IEEEauthorrefmark{2}}
\IEEEauthorblockA{
\IEEEauthorrefmark{1}Ericsson Research, Sweden\hspace{2mm}
\IEEEauthorrefmark{2}KTH Royal Institute of Technology, Sweden\\
Email: \{aditya.jolly, gabor.fodor\}@ericsson.com,~~~\{vitalyp, emilbjo\}@kth.se}\vspace{-7mm}}

\maketitle

\begin{abstract}
To achieve the desired coverage and capacity levels, future terahertz (THz) wireless systems are envisioned to utilize extremely large antenna arrays. At THz frequencies, the combination of short wavelengths and large array apertures often makes many of the conventional far-field assumptions invalid in practice. As a result, many UEs operate in the radiative near-field zone, where novel near-field beam synthesis methods become viable. This paper studies phase-only Bessel-like near-field beam configurations for downlink THz multiple-input multiple-output links under imperfect UE location knowledge. We first formulate a spectral efficiency maximization problem with respect to the ``Bessel cone angle''. We then derive \emph{low-complexity closed-form} approximations for the optimal Bessel beam configuration for: (i)~deterministic UE location; (ii)~Gaussian and (iii)~uniform error in the UE location. Finally, through extensive simulations across multiple signal frequencies, UE locations, and array sizes, we show that our proposed simple closed-form approximations closely match (under $0.1\%$ difference) the best performance achieved via exhaustive search, while simultaneously reducing the configuration complexity down to as low as $O(1)$. \vspace{-1mm}
\end{abstract}

\acresetall

\section{Introduction}
\ac{thz} ($0.3$--$3$~THz) communications can theoretically support up to terabit-per-second data rates by leveraging the large available bandwidth~\cite{spec_manage}. However, severe propagation losses make practical mobile THz deployments challenging and necessitate the use of steerable high-gain antennas (e.g., large-scale antenna arrays) to maintain a sufficient link budget~\cite{THz_problems}. As wavelengths shrink and large-scale arrays become more prevalent, the so-called radiative near-field zone can extend to several hundred meters~\cite{review_THz}. As a result, a substantial fraction of \acp{ue} may operate in this zone, where conventional far-field beamforming based on plane-wave assumptions can incur significant performance losses~\cite{Primer_Emil}.

Motivated by this, recent work has increasingly focused on near-field-specific beams, such as \emph{beam focusing}~\cite{review_THz}, which concentrates array energy at a spatial focal point rather than only an angular direction~\cite{BF_NF}. 
Prior works have quantified its achievable gains, developed phase-optimization algorithms and near-field codebooks~\cite{BF_NF,BF_BB_Channel_meas,Primer_Emil,BF_DA,BF_Opt,Codebook-design_BF}, among others. 
However, beyond single-point energy concentration offered by beam focusing, near-field operation also enables a much broader class of beams with distinct propagation characteristics, including Airy beams~\cite{airy}, Weber/Mathieu beams~\cite{W_and_M}, Hermite--Gaussian beams~\cite{H_Gauss}, and vortex beams~\cite{Vortex_beam_alignment}, among others.

In this context, \emph{Bessel-like beams}~\cite{BB_Gen} (hereafter \emph{Bessel beams}) have attracted significant attention for applications in mobile \ac{thz} links. This is primarily due to two key properties of Bessel beams. First, their \emph{``non-diffracting behavior''}~\cite{BB_Old}, meaning that they sustain axial power over a finite range and can therefore provide robustness to \ac{ue} location uncertainty~\cite{BB_6G,spec_manage,ULA_NF_Beams_limit}, and second, their \emph{``self-healing capability''}~\cite{wavefront_eng}, meaning that they can reconstruct the signal after partial link occlusion and thereby improve resilience to small-scale blockage~\cite{spec_manage,ULA_NF_Beams_limit,opt_mrt_BB}.

Recent work spans efficient beam generation and hardware-oriented synthesis to experimental demonstrations showing that Bessel beam-based transmission can enhance \ac{se} in ultra-wideband \ac{thz} systems~\cite{BB_Gen,Nature_Vitaly_bessel_beams,BB_NF_Link}.

However, the use of Bessel beams for practical mobile \ac{thz} communications raises an important question: \emph{``How to best configure a \ac{thz} Bessel beam for a given setup?''}. Unlike configuring far-field beamforming, which is mainly determined by the angle towards the \ac{ue} in line-of-sight, and near-field beam focusing, which depends on both the angle and the distance to the \ac{ue}  when known, Bessel beams introduce an additional design parameter: the ``Bessel cone angle''.

This makes the overall task of finding the optimal beam configuration non-trivial. Prior works typically either: (i)~adopted certain heuristic parameter choices~\cite{Nature_Vitaly_bessel_beams,BB_6G}, which may be suboptimal, or (ii)~performed exhaustive search over available configurations~\cite{BB_NF_Link}, which is computationally expensive and may lead to control signaling overheads between the \ac{ap} and the \ac{ue}. Hence, neither of the two approaches is suitable for prospective practical \ac{thz} deployments, motivating the need for \emph{an efficient yet low-complexity approach to optimize the \ac{thz} Bessel beam configuration}.

The problem is further complicated by the fact that, in practice, \ac{ue} location estimates are typically imperfect~\cite{BF_Opt}. Hence, the optimal Bessel beam configuration depends jointly on the \emph{estimated} \ac{ue} location, the uncertainty distribution, and the performance metric of interest (e.g., received power, \ac{snr}, or \ac{se}), making real-time adaptation with existing approaches challenging. Latest prior studies in this field (e.g.,~\cite{ULA_NF_Beams_limit} alongside other related works) derive certain boundaries for Bessel beam configuration and performance limits in different conditions for linear arrays under perfect knowledge of the \ac{ue} location. \emph{Still, to the best of the authors' knowledge, low-complexity techniques for optimizing Bessel beam configuration with realistic planar arrays and under given \ac{ue} location uncertainty remain under-explored.}

In this paper, we address this gap by deriving \emph{approximate \underline{closed-form} expressions that map near-optimal Bessel beam configurations to available \ac{ue} location statistics}. We particularly optimize the \ac{thz} Bessel beam configuration such that the \emph{expected} \ac{se} (and, consequently, capacity) of the AP--UE \ac{thz} wireless link is maximized. Through extensive computer simulations, we demonstrate that the proposed expressions closely track the best \ac{se} obtained via exhaustive search of available configurations, across a wide range of array sizes, frequencies, \ac{ue} locations, and uncertainty regimes. These closed-form approximations facilitate: (i)~low-complex (down to $\mathcal{O}(1)$) yet (ii)~efficient (negligible difference in \ac{se} compared to exhaustive search) Bessel beam configuration for the design and evaluation of future near-field \ac{thz} systems.

\section{System Model and Problem Formulation}{\label{sec:systmodel}}
We consider a single-user downlink near-field \ac{thz} \ac{mimo} system, where an \ac{ap} with an $N_{\rm AP}\!\times\! N_{\rm AP}$ \ac{upa} serves a \ac{ue} with an $N_{\rm UE}\!\times\! N_{\rm UE}$ \ac{upa}. Each antenna element is modeled as an ideal isotropic, linearly polarized point source (i.e., polarization mismatch and cross-polarization are neglected). Both \acp{upa} use $\lambda/2$ element spacing, and mutual coupling is neglected. The \ac{ap} transmits with total power $P_{\rm tx}$ (mW) and applies a conic phase profile to generate a Bessel beam (see Fig.~\ref{fig:Bessel Phase}), while the \ac{ue} performs coherent receive combining 
~\cite{MRC}.

Both \acp{upa} are parallel to the $xy$-plane, with the \ac{ap}'s \ac{upa} centered at the origin. The \ac{ue} is located on the \ac{ap} boresight and displaced only along the $z$-axis. We study three \ac{ue} location scenarios: (i) deterministic location, (ii) Gaussian-distributed errors along the $z$-axis, and (iii) uniformly-distributed errors along the $z$-axis, as illustrated in Fig.~\ref{fig:Comms-setup}.

\subsection{Near-field channel model}
We adopt a line-of-sight (LoS), non-uniform spherical-wave near-field channel model~\cite{Vitaly_SPAWC}. We operate in the radiative near-field region of the array such that $d_{\rm RA}< z \leq d_{\rm FA}$. Here, $d_{\rm RA}$ and $d_{\rm FA}$ denote the standard reactive and radiative near-field boundary, respectively~\cite{dfa}. The channel coefficient between the $(i,j)$-th \ac{ap} antenna element and the $(a,b)$-th \ac{ue} antenna element is denoted by $h_{(a,b)\rightarrow(i,j)}$, where $(i,j)$ and $(a,b)$ denote the row and column indices of the antenna elements on the \ac{ap} and \ac{ue} \acp{upa}, respectively, and $i,j \in \{1,\dots,N_{\rm AP}\}$ and $a,b \in \{1,\dots,N_{\rm UE}\}$. Then $h_{(a,b)\rightarrow(i,j)}$ is given by
\begin{equation}
h_{(a,b) \rightarrow (i,j)}
=
\frac{\lambda}{4\pi r_{(i,j)\rightarrow(a,b)}}
\exp\!\left(-j \frac{2\pi}{\lambda} r_{(i,j)\rightarrow(a,b)}\right),
\label{eq:channel}
\end{equation}
 where the scalar $r_{(i,j)\rightarrow(a,b)} \in \mathbb{R}^+$ denotes the propagation distance between the corresponding antenna elements.

\begin{figure}[t!]
\centering
\subfloat[System model with \ac{ue} mean location ($\mu$), and error distribution ($\varepsilon_{z})$]{%
\includegraphics[width=0.95\columnwidth]{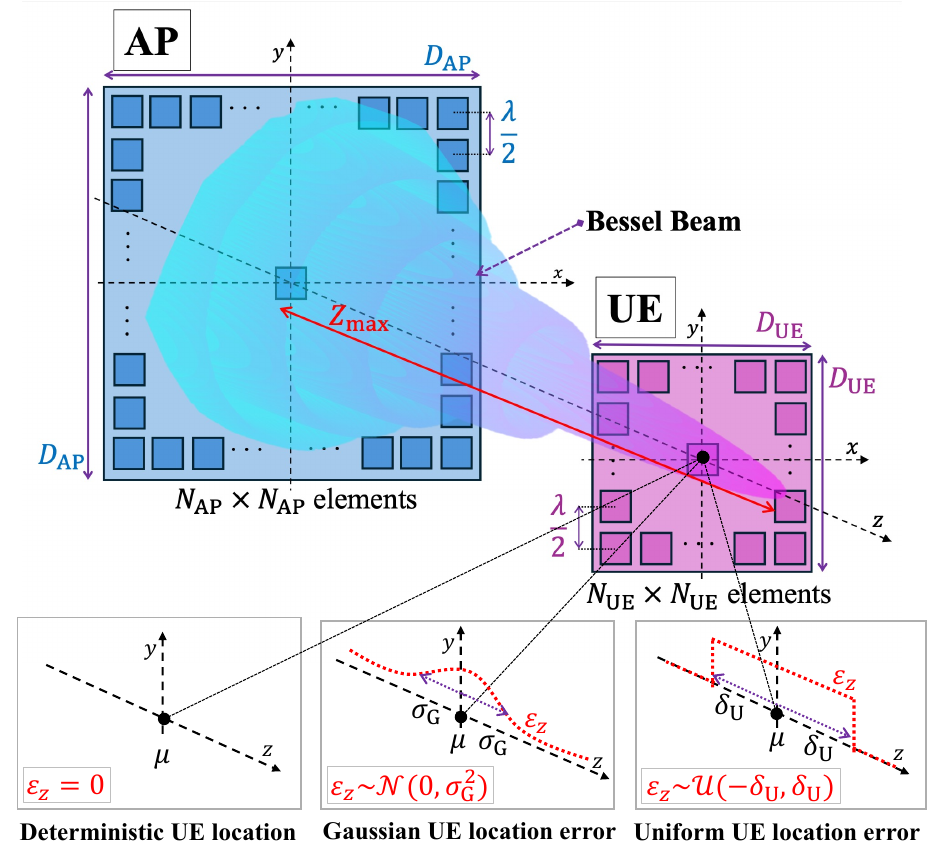}
\label{fig:Comms-setup}}
\\[0.1mm]
\subfloat[Bessel beam field intensity generated by conic phase profile ($\phi^{\rm BB}_{i,j}$)]{%
\includegraphics[width=0.95\columnwidth]{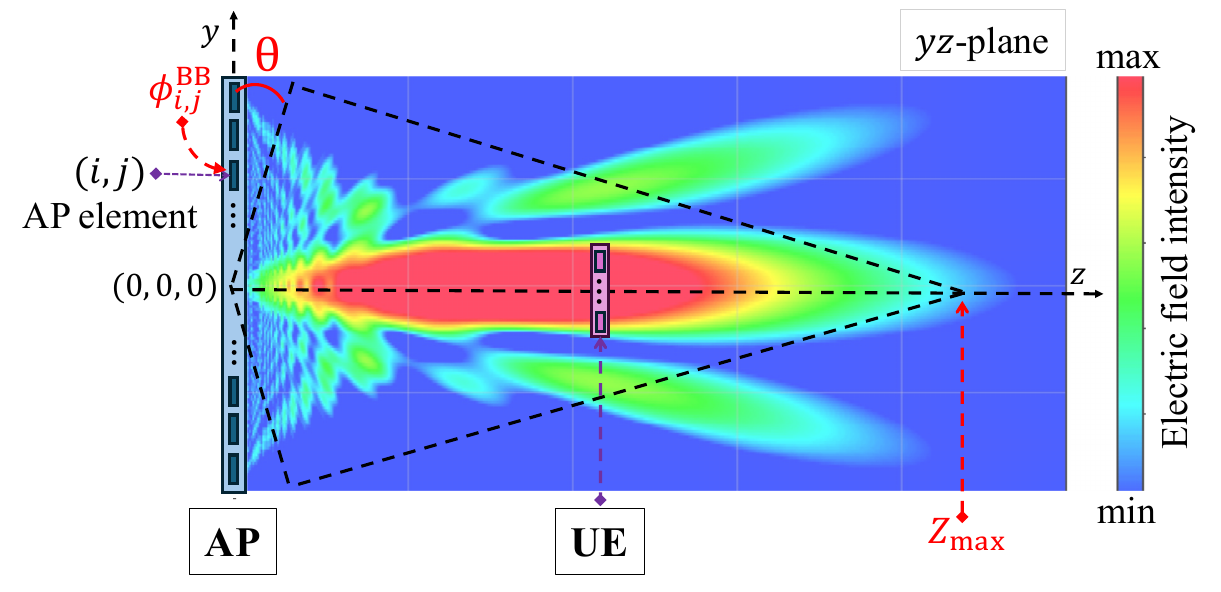}
\label{fig:Bessel Phase}}
\caption{Downlink near-field \ac{thz} \ac{ap}–\ac{ue} \ac{mimo} system utilizing a Bessel beam generated by a conic phase profile ($\phi^{\rm BB}_{i,j}$) with ``cone angle'' $\theta$.}
\label{fig:systmodel}
\vspace{-14pt}
\end{figure}

Let $\mu\in\mathbb{R}_{+}$ denote the estimated \ac{ue} location along the $z$-axis obtained from any positioning technique. Then the true distance between the antenna elements lying at the center of the \ac{ap} and \ac{ue} array is modeled as $p = \mu + \varepsilon_z$,
where $\varepsilon_z$ is an additive error assumed identical for all \ac{ue} antennas and confined to the $z$-axis (to isolate range uncertainty while excluding lateral misalignment to study the relationship of distance errors and near-field beam configuration).    
The resulting propagation distance is given by
\begin{equation}
r_{(i,j)\rightarrow(a,b)}
=\sqrt{(x_{a,b}-x_{i,j})^2+(y_{a,b}-y_{i,j})^2 + p^2},
\end{equation}
where $x_{i,j} \triangleq (i-(N_{\rm AP}+1)/2)\frac{\lambda}{2},\ y_{i,j}\triangleq(j-(N_{\rm AP}+1)/2)\frac{\lambda}{2}$ and $x_{a,b}\triangleq (a-(N_{\rm UE}+1)/2)\frac{\lambda}{2},\ y_{a,b}\triangleq(b-(N_{\rm UE}+1)/2)\frac{\lambda}{2}$.

For the scenario with a deterministic \ac{ue} location, the \ac{ue} is located at $\mu$ with no positioning error, i.e.,
$\varepsilon_z = 0$. 
Under the uniform positioning error model, the \ac{ue}'s location error is modeled as $\varepsilon_z \sim \mathcal{U}\!\left[-\delta_{\rm U},\,\delta_{\rm U}\right]$,
where the variance of $\varepsilon_{z}$ is $\sigma^{2}_{\rm U} = \delta_{\rm U}^2/3$. 
While, for the Gaussian positioning error model, the \ac{ue}'s location error is modeled as $
\boldsymbol{\varepsilon}_z \sim \mathcal{N}\!\left(0,\,\sigma_{\rm G}^2\right)$.
In both cases, the \ac{ue} location is symmetrically distributed around $\mu$ (i.e., $\varepsilon_z$ is zero-mean), as shown in Fig.~\ref{fig:Comms-setup}.

\subsection{Bessel beam generation}
To generate a Bessel beam using a phase-only technique, we apply a phase difference of $\mathbf{\Phi}^{\rm BB} \in \mathbb{R}^{N_{\rm AP} \times N_{\rm AP}}$ to the \ac{ap} \ac{upa}~\cite{BB_6G}. The per-element phase difference applied is denoted by $\phi^{\rm BB}_{i,j}$ (assumed to be continuous), and defined as
\begin{equation}
\phi^{\rm BB}_{i,j} = \frac{2\pi}{\lambda}\Big(\sqrt{x^2_{i,j} + y_{i,j}^2}\Big)\sin\theta.
\label{eq:phase_diff_ele}
\end{equation}
This phase-based technique applies a conic phase profile with ``cone angle'' $\theta$ at the transmitting aperture to generate a Bessel beam (see Fig.~\ref{fig:Bessel Phase}). Its non-diffracting behavior is characterized by $Z_{\rm max}$, given by $Z_{\rm max}=R/\tan\theta$, where $R$ is the effective aperture radius. We adopt the circumscribed radius $R=D_{\rm AP}/\sqrt{2}$, with $D_{\rm AP}=(N_{\rm AP}-1)\lambda/2$ denoting the side length of the \ac{ap} \ac{upa}. For a fixed frequency and aperture, $\theta$ fully determines the Bessel beam's characteristics.
\subsection{Problem formulation}{\label{sec:prob_form}}
We formulate Bessel beam configuration as the maximization of the expected \ac{se} under the given location statistics.  
With matched-filtering, the post-combining received signal power equals the squared $\ell_2$-norm of the per-antenna received signal:
\begin{equation}
P_{\rm UE} =
\sum_{a=1}^{N_{\rm UE}}\sum_{b=1}^{N_{\rm UE}}
\left\|
\sum_{i=1}^{N_{\rm AP}}\sum_{j=1}^{N_{\rm AP}}
\frac{\sqrt{P_{\rm tx}}}{N_{\rm AP}}\,
h_{(a,b) \rightarrow (i,j)}
e^{j\phi^{\text{BB}}_{i,j}}
\right\|^2.
\label{eq:Tot_Pow}
\end{equation}
Let $n_{\rm F}$ denote the post-combining noise power. The resulting \ac{snr} is $P_{\rm UE}/n_{\rm F}$, and the \ac{se} becomes $S=\log_2\!\big(1+\rm P_{\rm UE}/n_{\rm F}\big)$.
Our design objective is to choose $\theta$ that maximizes the expected \ac{se} under the \ac{ue} location statistics:
\begin{equation}
\theta^*=\arg\max_{\theta}\; \mathbb{E}\!\left[S(\theta)\right].
\label{eq:opt_problem}
\end{equation}
The problem in \eqref{eq:opt_problem} can be solved via exhaustive search over a discrete set of candidate ``cone angle'' $\theta$, yielding a benchmark for each \ac{ue} location scenario. However, this is computationally expensive. Hence, in the following section, we develop closed-form approximations to solve \eqref{eq:opt_problem}, under the considered \ac{ue} location scenarios.

\section{Optimal Bessel Beam Configuration}{\label{sec:bb_opt}}
In this section, we present the exhaustive search benchmark and derive closed-form Bessel beam configuration rules for three \ac{ue} location scenarios: deterministic location, Gaussian uncertainty, and uniform uncertainty. We first treat the deterministic \ac{ue} case using a geometric scaling argument, and validate the scaling law by quantifying its dependence on frequency and array size. We then extend the solution to the Gaussian and uniform uncertainty cases via first-order corrections and calibrate the associated correction factors.
\subsection{Exhaustive search benchmark}
 For each \ac{ue} operating point $(\mu,\sigma)$, with $\sigma\in\{0,\sigma_{\rm G},\sigma_{\rm U}\}$ denoting deterministic location, Gaussian, and uniform location uncertainty, respectively, we evaluate a discrete candidate set $\Theta\in [\theta_{\min}, \theta_{\max}]$. Since $\theta$ is the only design variable, \eqref{eq:opt_problem} is solved via a one-dimensional grid search. For each $\theta\in\Theta$, the \ac{se} is computed over a discrete axial grid $z\in[z_{\min},z_{\max}]$, and the expected \ac{se} is obtained by approximating the expectation via numerical averaging over the location distribution centered at $\mu$. The exhaustive search baseline is
\begin{equation}
\theta^*(\mu,\sigma)
=\arg\max_{\theta\in\Theta}\ 
\mathbb{E}_{z\sim \mathcal{L}(\mu,\sigma)}\!\left[S(\theta;z)\right],
\label{eq:exh_search}
\end{equation}
where $\mathcal{L}(\mu,\sigma)$ denotes the corresponding location distribution (degenerate for $\sigma=0$).
The cost of \eqref{eq:exh_search} is high because each evaluation of $S(\theta;z)$ requires a full spherical-wave near-field channel computation at distance $z$, with distance-dependent coefficients. Repeating this over many $\theta\in\Theta$ and $z$ samples to numerically average over $\mathcal{L}(\mu,\sigma)$ makes real-time operation prohibitive. In our implementation, this burden is partially mitigated using a GPU-accelerated, vectorized PyTorch in-house simulator with optimized precision and tiled processing.

\subsection{Closed-form approximation: Deterministic \ac{ue} location}\label{sec:axial_peak}
\begin{figure}[tbp!]
    \centering
	\includegraphics[width=0.99\columnwidth]{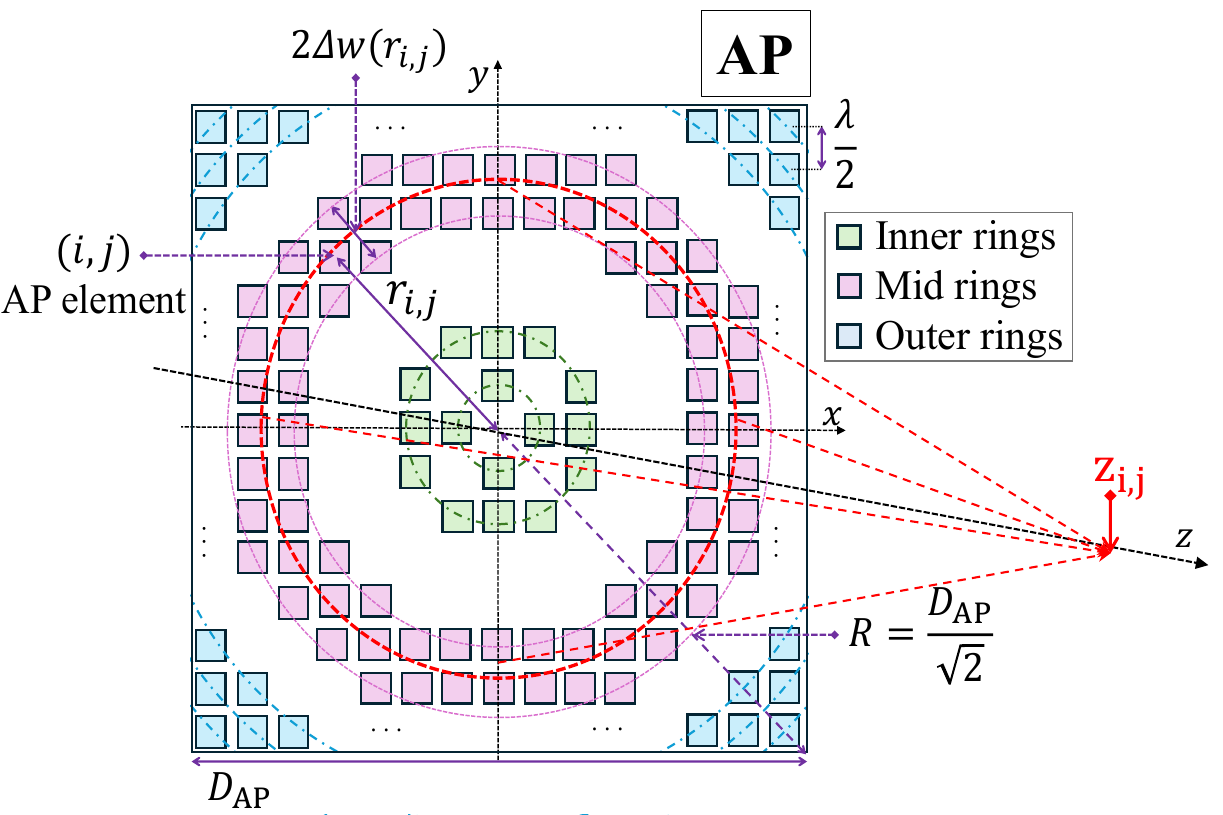}
	\caption{\ac{ap} antenna elements within an annular band of radius $r_{i,j}$ and width $2\Delta w(r_{i,j})$, which add quasi-coherently at $z_{i,j}$. Inner and mid rings (green/red) are untruncated, and their element count grows with radius, whereas near-boundary rings (blue) are truncated and contain fewer elements.}
	\label{fig:illustration}
    \vspace{-14pt}
\end{figure}
We first consider a single-antenna \ac{ue} and perfect location knowledge (deterministic) at the transmitter. To reveal the geometry governing the axial power profile, we apply a stationary-phase approximation to the overall phase term in \eqref{eq:Tot_Pow}. That is, we assume that the dominant contribution at a given $z$ comes from antenna elements whose phase derivative is \emph{approximately} zero. 

For an \ac{ap} antenna element $(i,j)$ with radial coordinate $r_{i,j}=\sqrt{x_{i,j}^2+y_{i,j}^2}$, this condition yields the constructive (stationary) axial distance $z_{i,j}=r_{i,j}/\tan\theta,\ \forall\, i,j$.
Consequently, the discrete square \ac{upa} can be interpreted \emph{approximately} as a set of ``rings'': elements with similar $r_{i,j}$ contribute quasi-coherently around the corresponding $z_{i,j}$ (see Fig.~\ref{fig:illustration}).
\begin{figure*}[t!]
\centering

\begin{minipage}[t]{0.66\textwidth}
\centering

\subfloat[Effect of aperture and frequency on \eqref{eq:final_result}]{%
  \includegraphics[width=0.475\linewidth]{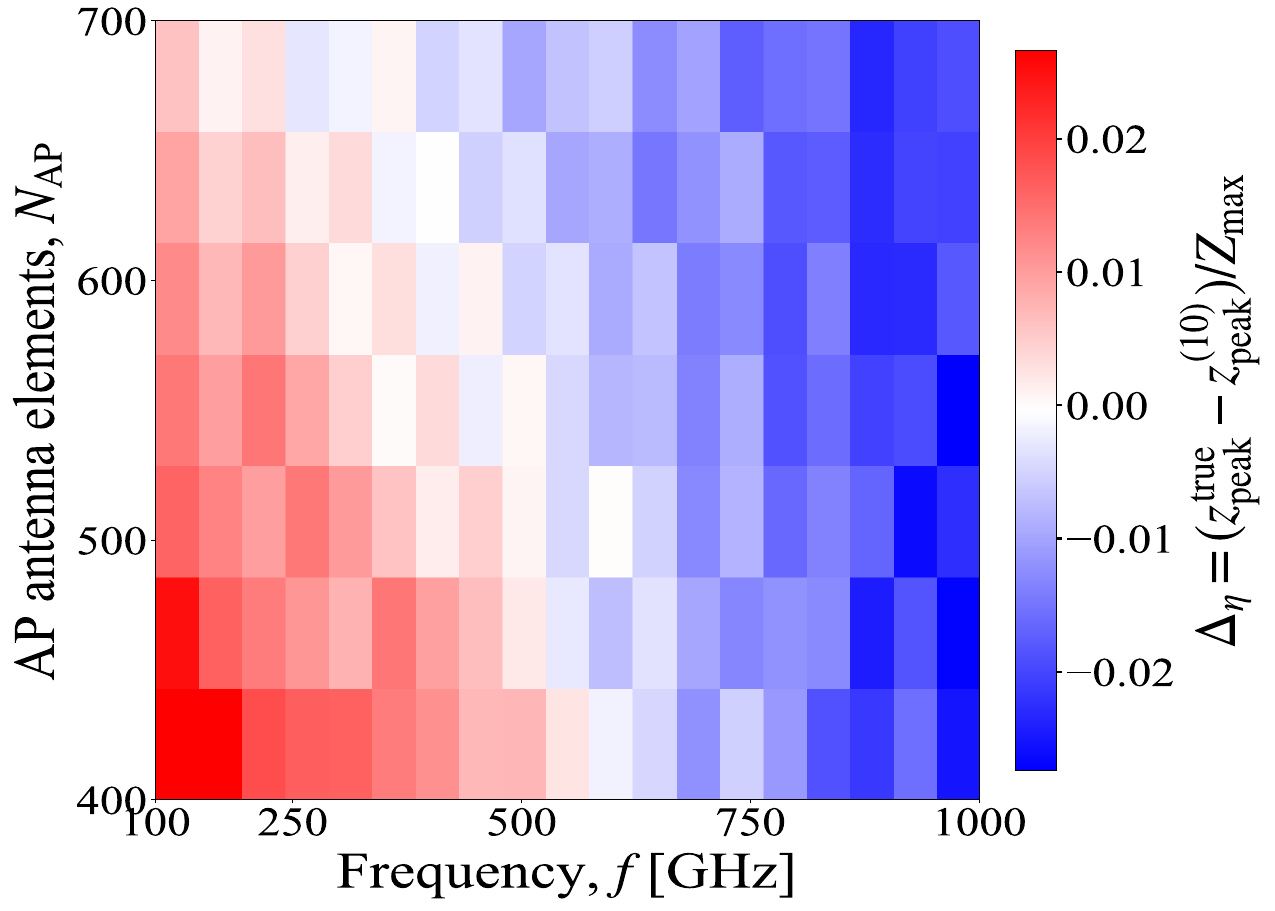}%
  \label{fig:result_eta}
}\hfill
\subfloat[Effect of aperture and frequency on $S^{\rm CF}_{\rm D}$, \eqref{eq:perfectue}]{%
  \includegraphics[width=0.475\linewidth]{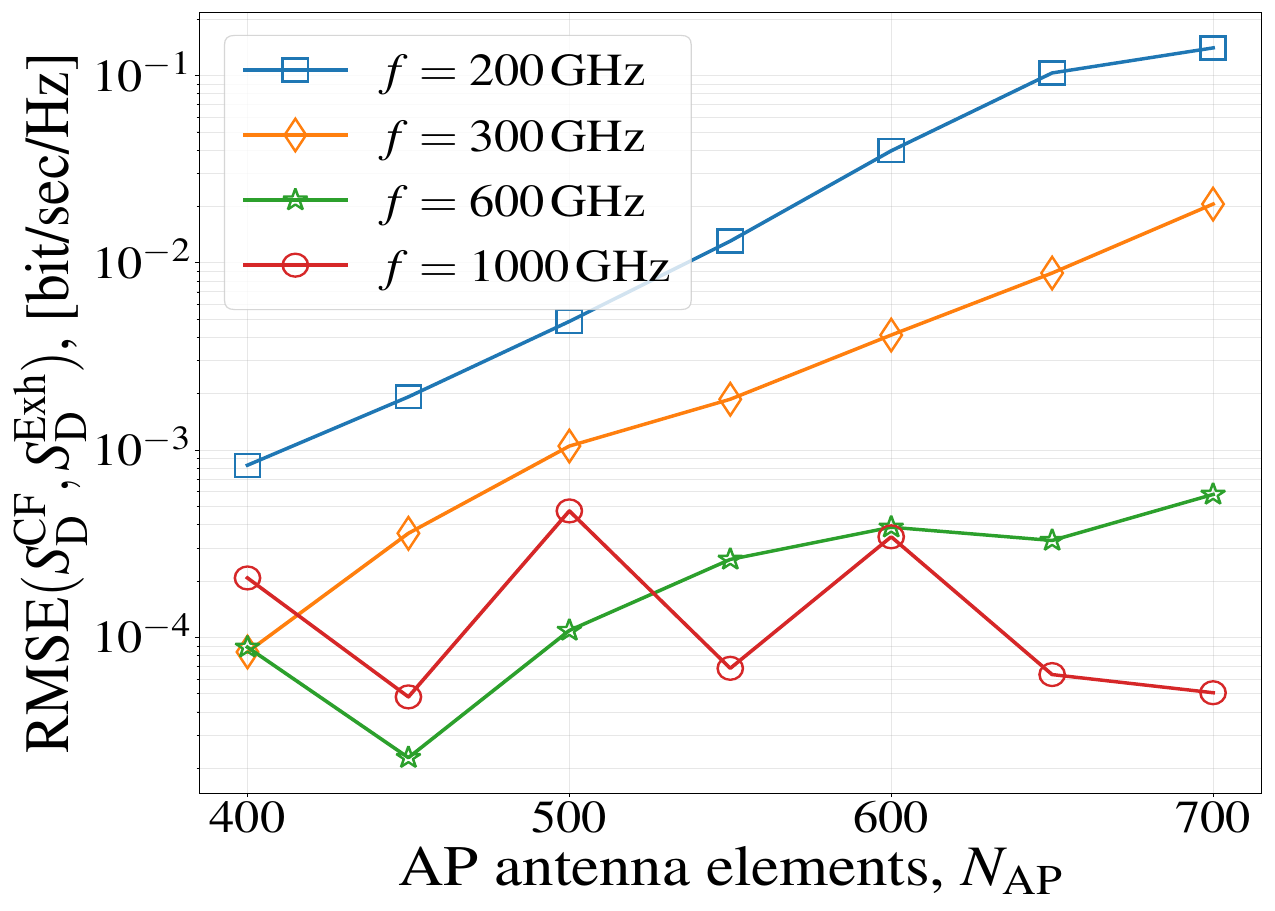}%
  \label{fig:P_new}
}

\caption{Validation for deterministic UE location case, (a) Heatmap of $\Delta_{\eta}=(z^{\rm true}_{\rm peak}-z^{\rm (10)}_{\rm peak})/Z_{\max}$ across $f$ and $N_{\rm AP}$, (b) RMSE between $S^{\rm Exh}_{\rm D}$ and $S^{\rm CF}_{D}$ (log-scaled) vs $N_{\rm AP}$, $\mu=10m$ and $N_{\rm UE}=1$.}
\label{fig:P_case_ab}

\end{minipage}
\hfill
\begin{minipage}[t]{0.32\textwidth}
\centering
\includegraphics[width=\linewidth]{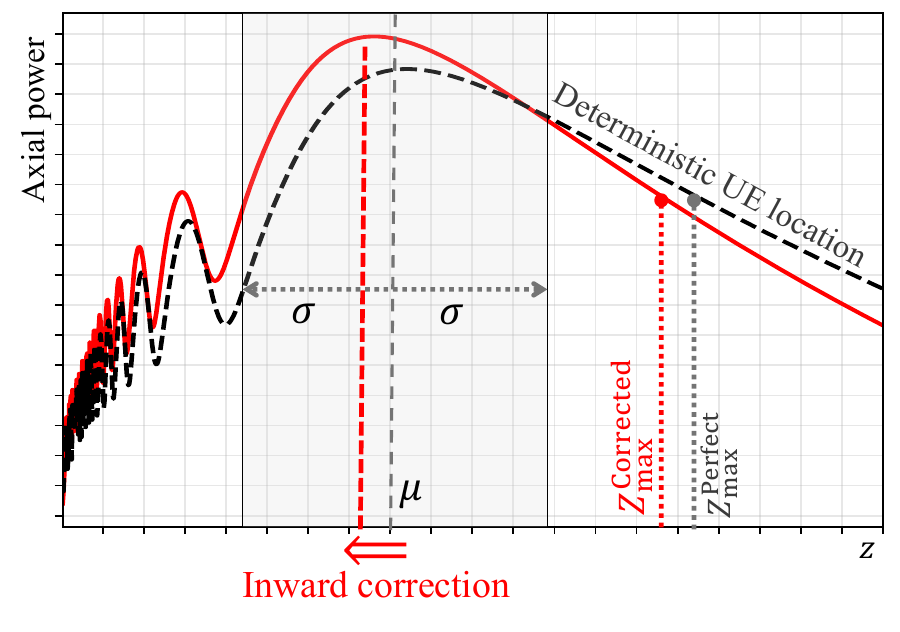}

\begingroup
\makeatletter
\def\@captype{figure}
\caption{Axial power profile of Bessel beams for deterministic (black) and uncertain (red) UE locations, depicting the \emph{peak-alignment} principle.}
\label{fig:peak_align}
\makeatother
\endgroup

\end{minipage}
\vspace{-10pt}
\end{figure*}

To obtain a tractable scaling law, we model the coherent set at radius $r_{i,j}$ as an annular band of thickness $2\Delta w (r_{i,j})$. We assume that over this narrow neighborhood, the propagation distance and phase curvature vary slowly, so the contributions remain nearly phase-aligned. For a finite square aperture with effective radius $R$, the band is truncated by the boundary, which we capture by
\begin{equation}
\Delta w(r_{i,j})=\min(r_{i,j},\,R-r_{i,j}).
\label{eq:ring_width}
\end{equation}
With sufficiently dense sampling and $d=\lambda/2$, the number of elements in this band is approximated via the area density as
\begin{equation}
N_{\text{ring}}(r_{i,j})\approx \frac{2\pi r_{i,j}\,\Delta w(r_{i,j})}{d^2}.
\label{eq:ring_count}
\end{equation}
Assuming coherent summation within the annular band, the field amplitude scales as $N_{\text{ring}}(r_{i,j})$ and the received power scales as $N_{\text{ring}}(r_{i,j})^2$, with an additional free-space decay $1/z^2$. Since the annular band at radius $r_{i,j}$ contributes primarily around $z = z_{i,j}=r_{i,j}/\tan\theta$, we evaluate the proxy \emph{along this mapping} by setting $r_{i,j}=z_{i,j}\tan\theta$, yielding
\begin{equation}
P_{\rm UE}(z_{i,j}) \propto \frac{\big(N_{\text{ring}}(z_{i,j}\tan\theta)\big)^2}{z_{i,j}^2}
 \propto \tan^2\theta\,(\Delta w(z_{i,j}\tan\theta))^2 .
\label{eq:power_model_z}
\end{equation}
Based on \eqref{eq:ring_width}, we notice that $\Delta w(z_{i,j}\tan\theta)$ grows linearly with $z_{i,j}\tan\theta$ until $z_{i,j}\tan\theta=R/2$ and then decreases linearly due to truncation (see Fig.~\ref{fig:illustration}). Therefore, $P_{\rm UE}(z_{i,j})$ increases for $z_{i,j}\le R/(2\tan\theta)$ and decreases for $z_{i,j}>R/(2\tan\theta)$, predicting the dominant axial peak at
\begin{equation}
z_{\text{peak}}=R/(2\tan\theta)=Z_{\max}/2.
\label{eq:final_result}
\end{equation}
We note that \eqref{eq:final_result} relies on a standard first-order approximation that captures the dominant scaling via coherent element density and $1/z^2$ spreading. However, \eqref{eq:final_result} is not intended to exactly characterize the peak location since contributions from elements outside the dominant annular region (depending on aperture and frequency) are neglected. This aligns with our goal of exposing the governing geometry while quantifying residual effects empirically. 

Using \eqref{eq:final_result}, we obtain a closed-form Bessel beam configuration for deterministic \ac{ue} location based on a \emph{peak-alignment principle}: choose the ``cone angle'' $\theta$ so that the dominant axial maximum of the finite-aperture beam aligns with the \ac{ue} location. Since the downlink \ac{snr} (and thus \ac{se}) is maximized when the received power peaks at the \ac{ue} position, this gives a simple, physically interpretable design rule (see Fig.~\ref{fig:peak_align}):
\begin{equation}
\mu = z_{\text{peak}} = Z_{\max}/2
\end{equation}
and yields the closed-form configuration
\begin{equation}
\theta^{*}_{\text{Deterministic}} =
\tan^{-1}\!\ \Bigg(\frac{D_{\rm AP}}{2\sqrt{2}\mu}\Bigg).
\label{eq:perfectue}
\end{equation}

Fig.~\ref{fig:P_case_ab} validates \eqref{eq:final_result}--\eqref{eq:perfectue} and quantifies how the axial peak depends on the aperture and frequency. Specifically, we sweep $f\in[100,1000]$~GHz (including sub-THz frequencies) with $N_{\rm UE}=1$, vary $N_{\rm AP}\in[400,700]$ (corresponding to practical apertures from $6$~cm at $1$~THz to $1.05$~m at $100$~GHz), and scan $\theta\in[0.1^{\circ},10^{\circ}]$. For each configuration, our near-field simulator computes the on-axis received power, from which we extract the true axial peak $z^{\rm true}_{\rm peak}$ and define the normalized ratio $\eta \triangleq z^{\rm true}_{\rm peak}/Z_{\max}$. Fig.~\ref{fig:result_eta} then reports $\Delta_{\eta}=\eta-z^{(10)}_{\rm peak}/Z_{\max}$ over $(f,N_{\rm AP})$, showing deviations tightly centered around zero (i.e., $\mathbb{E}[\eta]=0.497$ and $\mathbb{E}[\Delta_{\eta}]\!\approx\!5.3\times10^{-3}$) with only a weak residual dependence on $(f,N_{\rm AP})$.

Fig.~\ref{fig:P_new} quantifies the impact of this weak dependence on \ac{se} by reporting the \ac{rmse} (log-scaled $y$-axis) between the exhaustive-search benchmark ($S^{\rm Exh}_{\rm D}$)~\eqref{eq:exh_search} and the closed-form expression \ac{se} ($S^{\rm CF}_{\rm D}$) in~\eqref{eq:perfectue}. We vary the array size $N_{\rm AP}$ (x-axis) for $f\in\{200,300,600,1000\}$~GHz while fixing $\mu=10$~m and $N_{\rm UE}=1$. As seen in Fig.~\ref{fig:P_new}, the trends remain mild, with the maximum \ac{rmse} on the order of $10^{-1}$~bits/s/Hz. Overall, the peak-alignment rule $z_{\rm peak}\approx Z_{\max}/2$ is robust, and the residual aperture/frequency effects are small, so \eqref{eq:final_result}--\eqref{eq:perfectue} remain accurate in the deterministic-\ac{ue}-location case. Moreover, for an array-equipped \ac{ue} with coherent combining, \eqref{eq:perfectue} remains a valid first-order rule since $z_{\rm peak}$ is governed primarily by the transmit-side geometry.

\subsection{Closed-form approximation: Uncertain \ac{ue} location}\label{sec:uncertaincform}
Under location uncertainty, the peak-alignment principle is extended from a single-point target to a distribution-dependent target region. For Gaussian uncertainty, the ``cone angle'' is chosen to align the dominant axial maximum with the high-probability region of the \ac{ue} location distribution rather than the mean location alone. Exploiting the smooth axial variation of the Bessel beam power around $z_{\text{peak}}$, we model the resulting shift via a first-order correction, $\mu-\alpha\sigma_{\text{G}}= z_{\text{peak}} = Z_{\max}/2$.
Where $\alpha$ captures the effective inward correction induced by Gaussian uncertainty. 

The same principle applies to the uniform uncertainty case, yielding an analogous first-order correction, $\mu-\beta \sigma_{\rm U}
= z_{\text{peak}} = Z_{\max}/2$.
Where $\beta$ captures the corresponding effective inward correction under uniform uncertainty. As shown in Fig.~\ref{fig:peak_align}, the inward bias concentrates axial energy and moves the dominant peak toward the region maximizing the expected \ac{se}. Although $Z_{\max}$ decreases, the average \ac{se} improves under uncertainty since the near-peak power gain outweighs the roll-off beyond $Z_{\max}$. The coefficients $\alpha$ and $\beta$ thus quantify the strength of this uncertainty-induced inward shift.

We obtain $\alpha$ (Gaussian) and $\beta$ (uniform) by calibrating the closed-form rules against the exhaustive-search benchmark in~\eqref{eq:exh_search}. The exhaustive search is performed over a $\Theta\in [0.1^{\circ},10^{\circ}]$ with a granularity of $0.05^{\circ}$, and $z\in[d_{\rm RA},d_{FA}/5]$ with a granularity of $10$~cm. 

\begin{figure}[t!]
\centering
\subfloat[Gaussian error in \ac{ue} location]{%
\includegraphics[width=0.96\columnwidth]{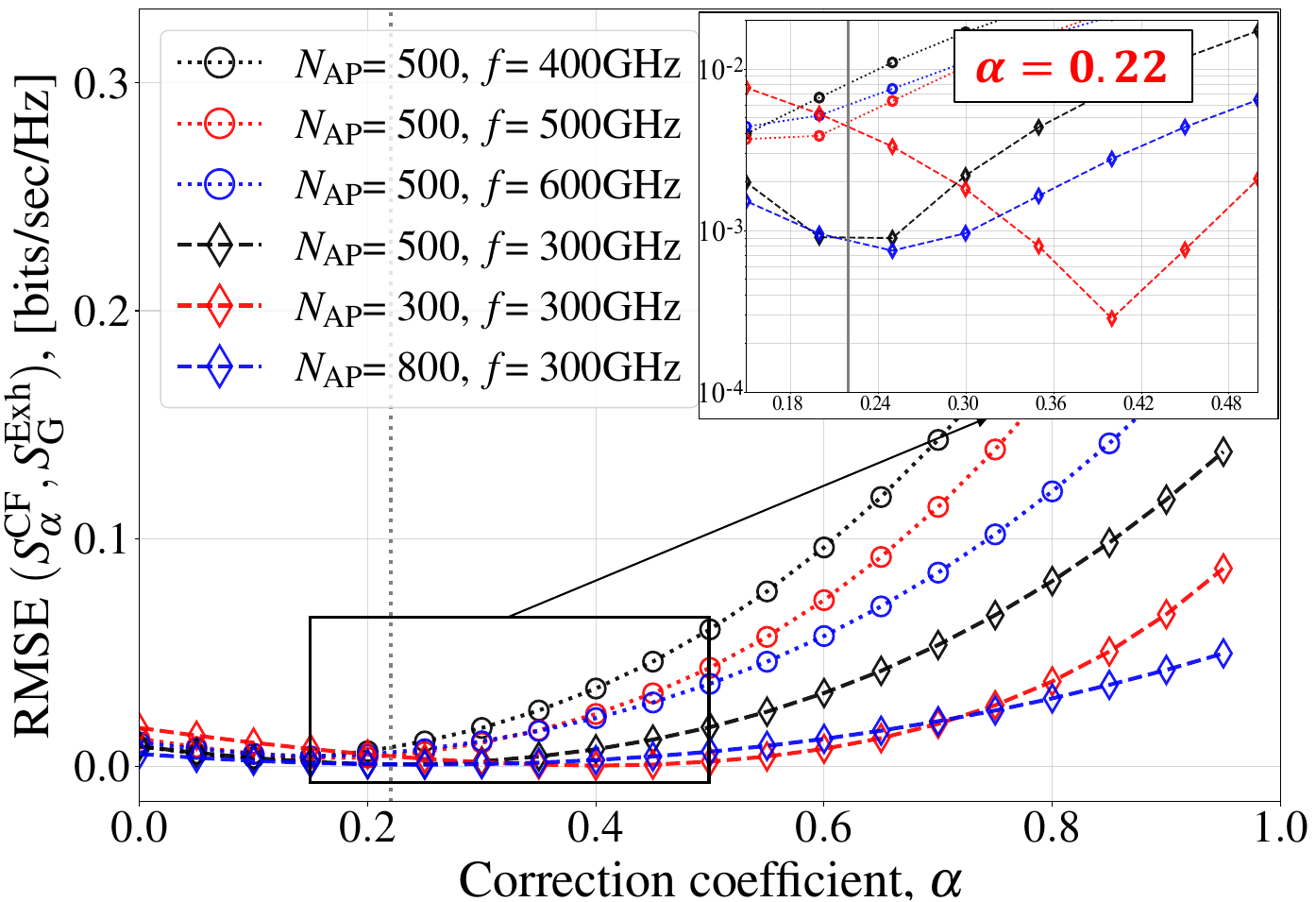}
\label{fig:G_alpha}}
\\[0.1mm]
\subfloat[Uniform error in \ac{ue} location]{%
\includegraphics[width=0.96\columnwidth]{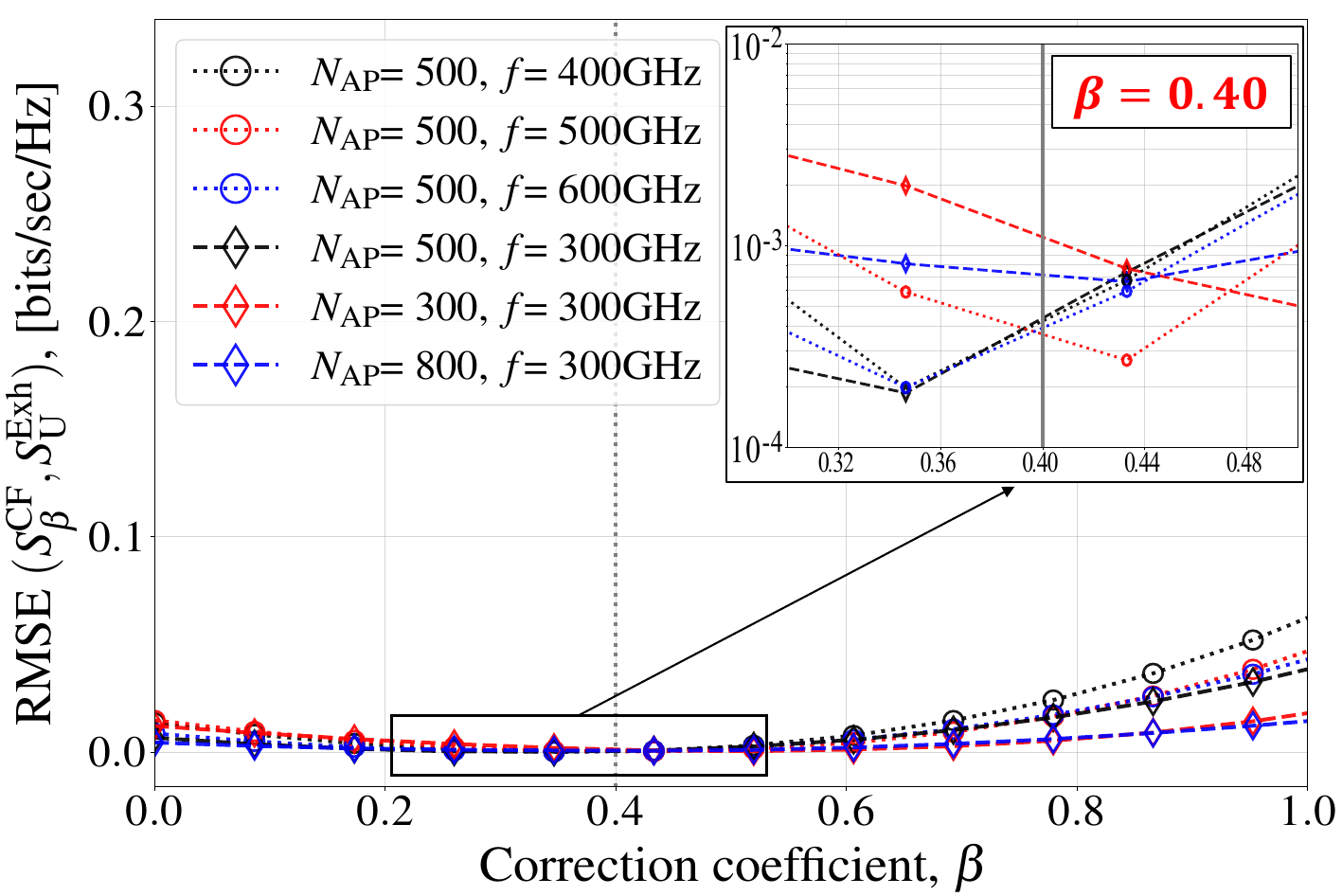}
\label{fig:U_beta}}
\caption{\ac{rmse} between $S^{\rm Exh}_{\rm G}$ and $S^{\rm CF}_{\alpha}$ for Gaussian error, and $S^{\rm Exh}_{\rm U}$ and $S^{\rm CF}_{\beta}$ for uniform error, averaged over (a) $[\mu,\ \sigma_{\rm G}]$, and (b) $[\mu,\ \sigma_{\rm U}]$.}
\label{fig:alpha}
\vspace{-17pt}
\end{figure}

For calibration, we sweep candidate coefficients from $\alpha,\beta\in[0.1,1]$ and, for each coefficient, compute the predicted ``cone angle'' and the resulting \emph{expected} \ac{se}, denoted by $S^{\rm CF}_{\alpha}(\mu,\sigma_{\rm G})$ for Gaussian errors and $S^{\rm CF}_{\beta}(\mu,\sigma_{\rm U})$ for uniform errors. We define $S^{\rm Exh}_{\rm G}(\mu,\sigma_{\rm G})$ and $S^{\rm Exh}_{\rm U}(\mu,\sigma_{\rm U})$ as the expected \ac{se} attained by \eqref{eq:exh_search} for Gaussian and uniform errors, respectively. In Fig.~\ref{fig:alpha}, we then measure the mismatch via \ac{rmse}, averaged over $(\mu,\sigma_{\rm G})$ and $(\mu,\sigma_{\rm U})$ grids with $\sigma_{\rm G}\in[1,5]$\,m and $\sigma_{\rm U}\in[0.57,2.89]$\,m.
For illustration, we present the results for three \ac{ap} sizes $N_{\mathrm{AP}}\in\{300,500,800\}$ at $f=300$\,GHz and three frequencies $f\in\{400,500,600\}$\,GHz at $N_{\mathrm{AP}}=500$.

Fig.~\ref{fig:alpha} reports the \ac{rmse} versus $\alpha$ (Gaussian) and $\beta$ (uniform); the insets zoom into the low-\ac{rmse} region and use a log-scale to resolve small differences among minima. The minima occur at slightly different values across configurations, but show no systematic trend with either $f$ or $N_{\mathrm{AP}}$. We set the coefficients to the mean of the per-configuration minimizers, yielding $\alpha=0.22$ and $\beta=0.40$, which provide near-minimal \ac{rmse} across the full test set. Hence, the proposed coefficients robustly link UE uncertainty statistics to the Bessel beam configuration. Substituting the values obtained for $\alpha$ and $\beta$ yields the following closed-form approximations
\begin{equation}
    \theta^{*}_{\text{Gaussian}} = \tan^{-1}\Bigg(\frac{D_{\rm AP}}{2\sqrt{2}\ (\mu - 0.22\ \sigma_{\mathrm{G}})}\Bigg),
    \label{eq:gauss}
\end{equation}
\begin{equation}
        \theta^{*}_{\text{Uniform}} = \tan^{-1}\Bigg(\frac{D_{\rm AP}}{2\sqrt{2}\ (\mu - 0.40\ \sigma_{\mathrm{U}})}\Bigg).
        \label{eq:uni}
\end{equation}
 With $(\alpha,\beta)$ fixed, \eqref{eq:gauss}--\eqref{eq:uni} fully specify the proposed closed-form configurations under Gaussian and uniform location errors, while \eqref{eq:perfectue} gives the corresponding closed-form approximation for deterministic \ac{ue} location. 

\begin{figure}[!b]
    \vspace{-5mm}
    \centering
	\includegraphics[width=0.96\columnwidth]{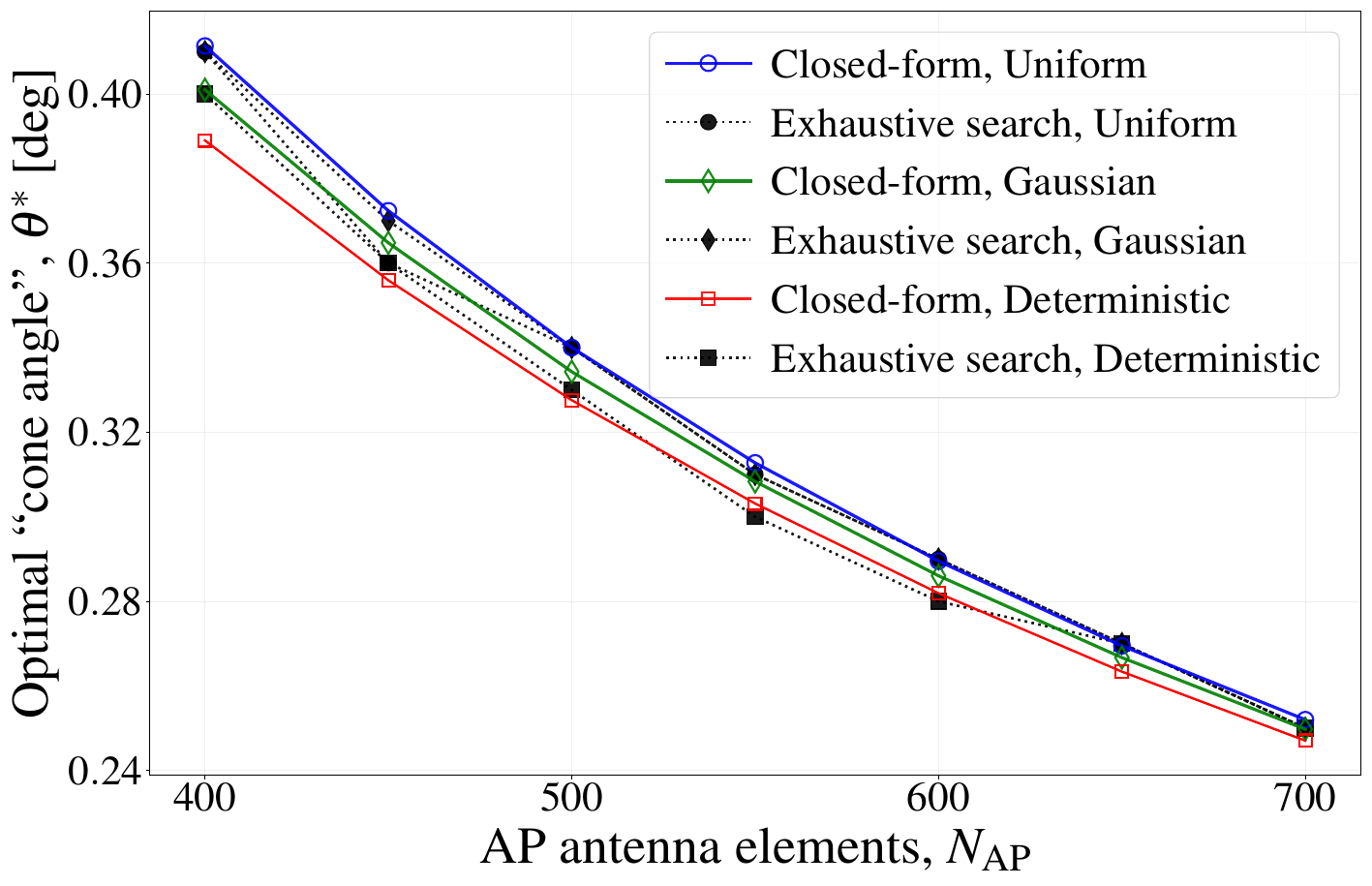}
    \vspace{-2mm}
	\caption{Optimal ``cone angle'' $\theta^*$ for closed-form vs exhaustive search.}
	\label{fig:theta_vs_nap}
\end{figure} 

\begin{figure*}[t!]
\centering
\subfloat[CDFs of $|\Delta S^{\%}_{\rm D}|$, deterministic location]{%
\includegraphics[width=0.66\columnwidth]{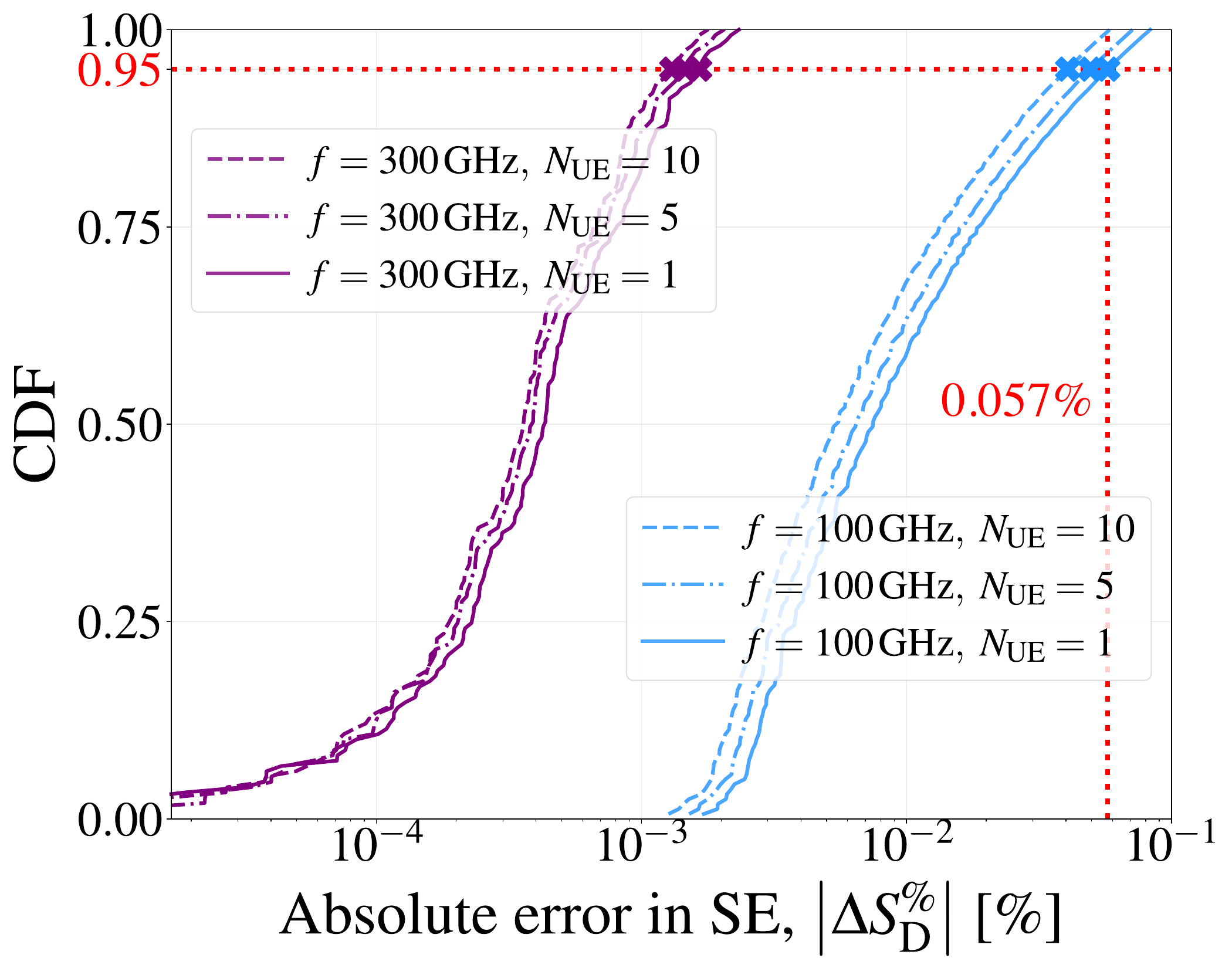}
\label{fig:P_CDF}}
\hfill
\subfloat[CDFs of $|\Delta S^{\%}_{\rm G}|$, Gaussian location error]{%
\includegraphics[width=0.66\columnwidth]{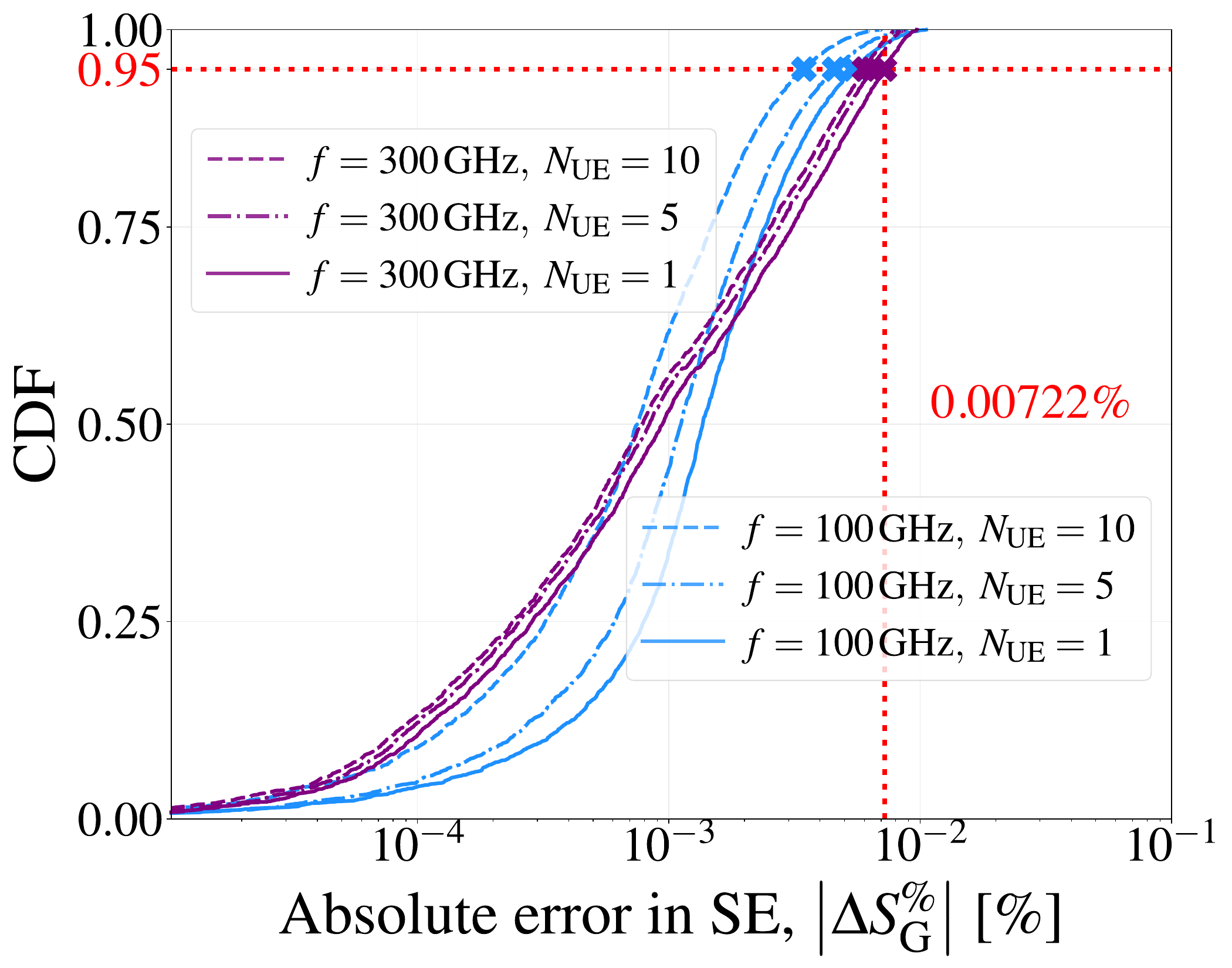}
\label{fig:U_CDF}}
\hfill
\subfloat[CDFs of $|\Delta S^{\%}_{\rm U}|$, uniform location error]{%
\includegraphics[width=0.66\columnwidth]{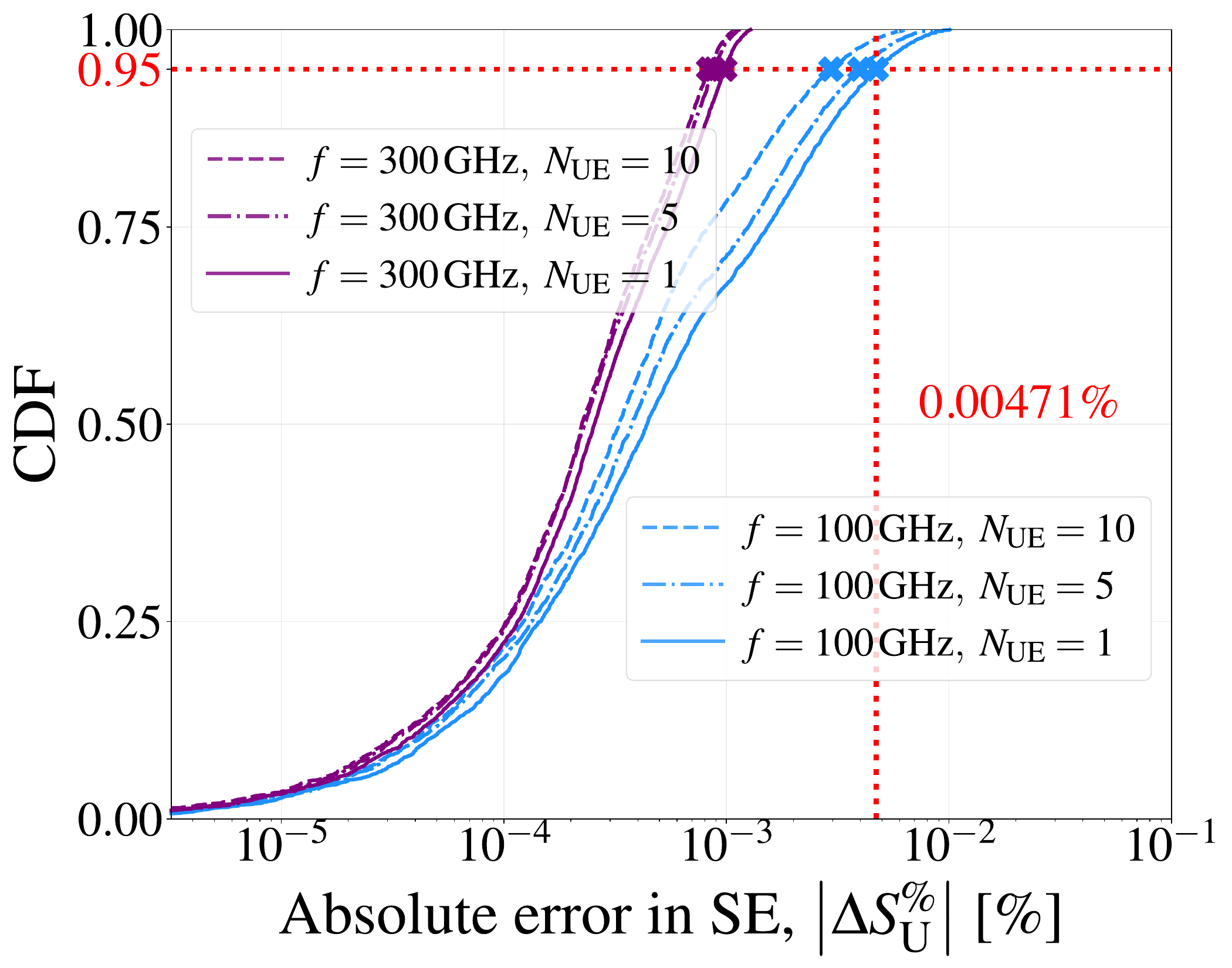}
\label{fig:G_CDF}}
\caption{CDFs of absolute percentage error in SE $|\Delta S^{\%}_{\rm Dist}|$ showing the impact of varying $N_{\rm UE}$ and $f$, on accuracy of $S^{\rm CF}_{\rm Dist}$, for three \ac{ue} location scenarios. }
\label{fig:CDF} 
\vspace{-5mm}
\end{figure*}

\begin{figure*}[t]
\centering
\subfloat[Effect of frequency on \ac{se}, deterministic case]{%
\includegraphics[width=0.66\columnwidth]{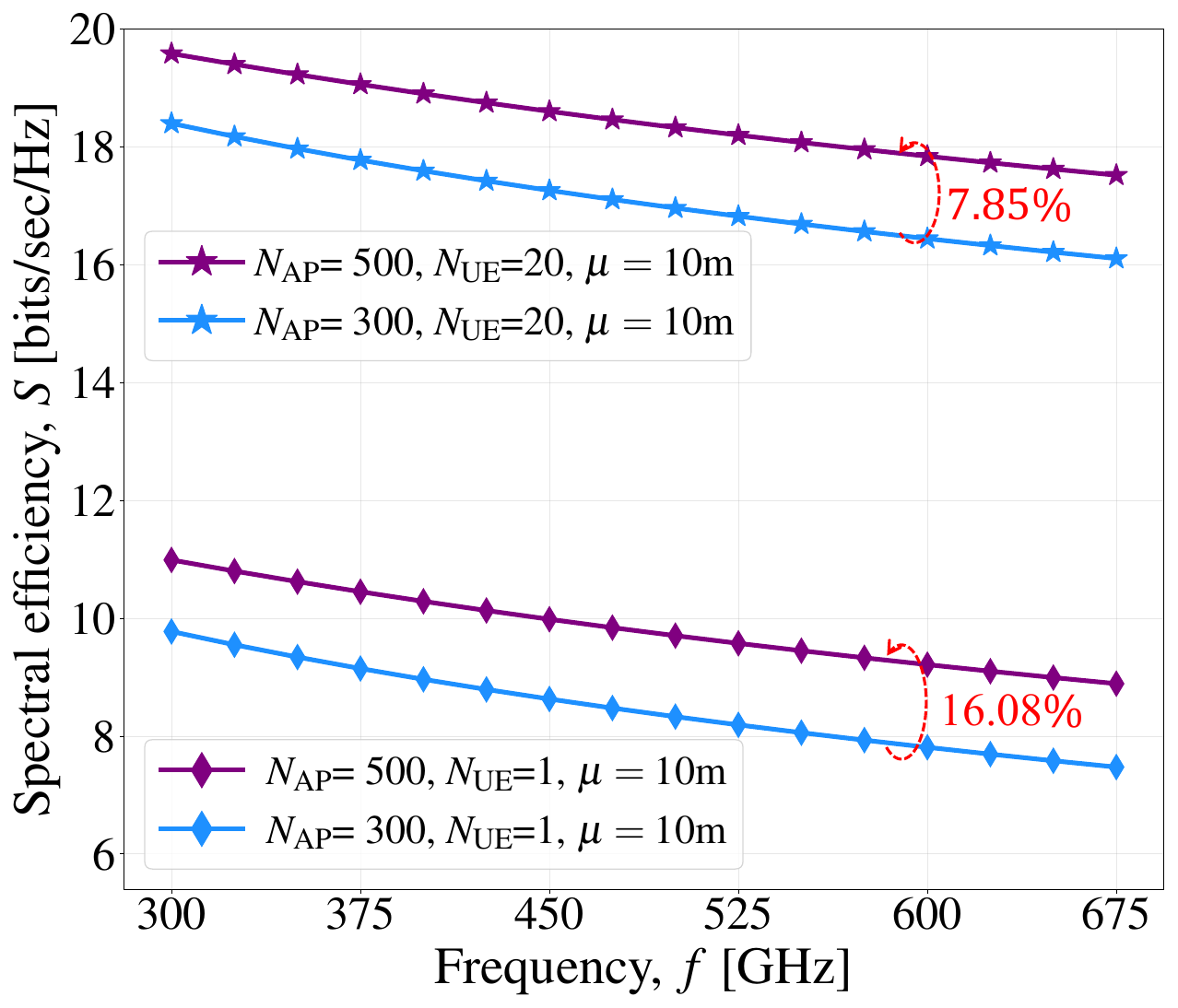}
\label{fig:SE}}
\hfill
\subfloat[Effect of UE mean location ($\mu$) on \ac{se}]{%
\includegraphics[width=0.66\columnwidth]{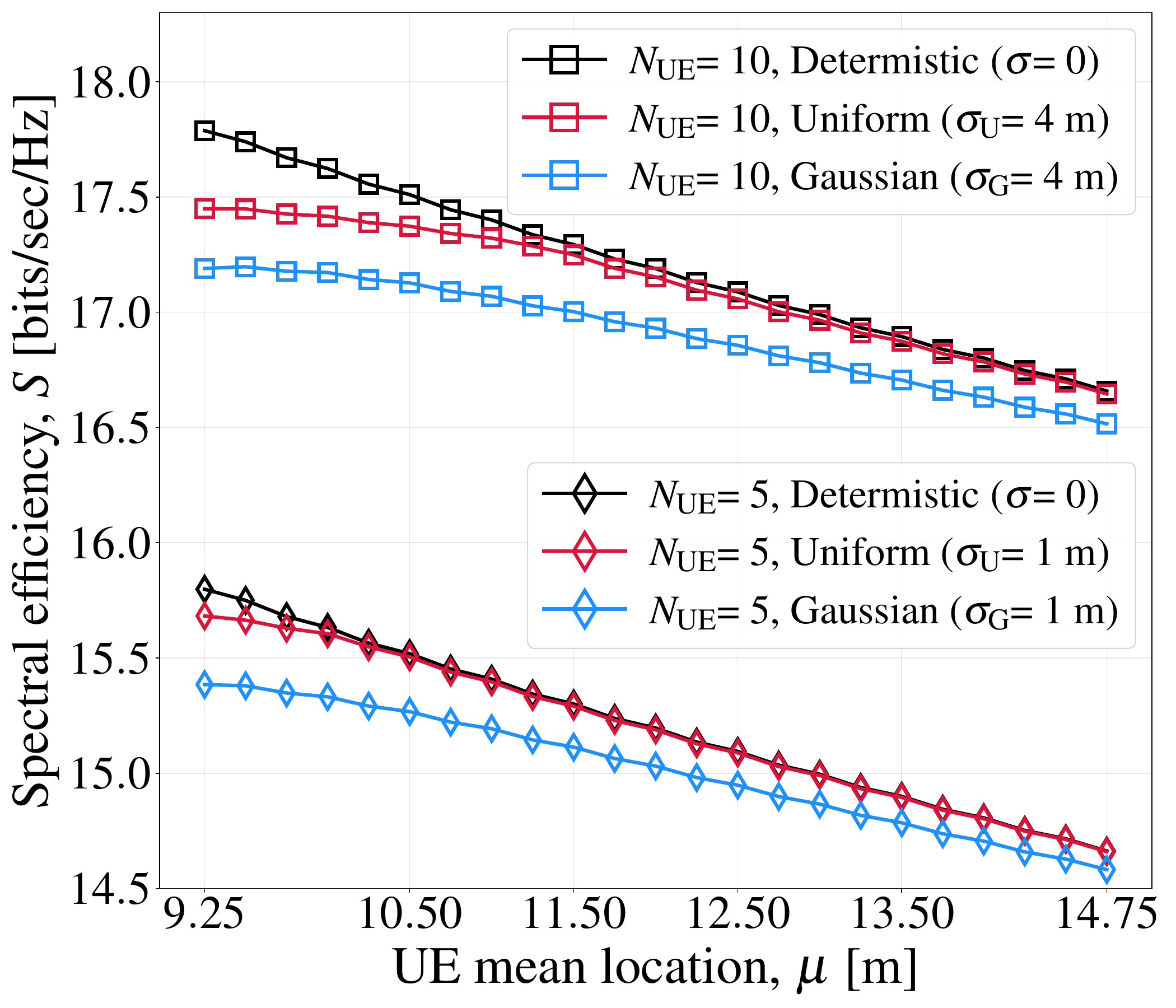}
\label{fig:SEvsU}}
\hfill
\subfloat[Effect of UE location uncertainty ($\sigma$) on \ac{se}]{%
\includegraphics[width=0.66\columnwidth]{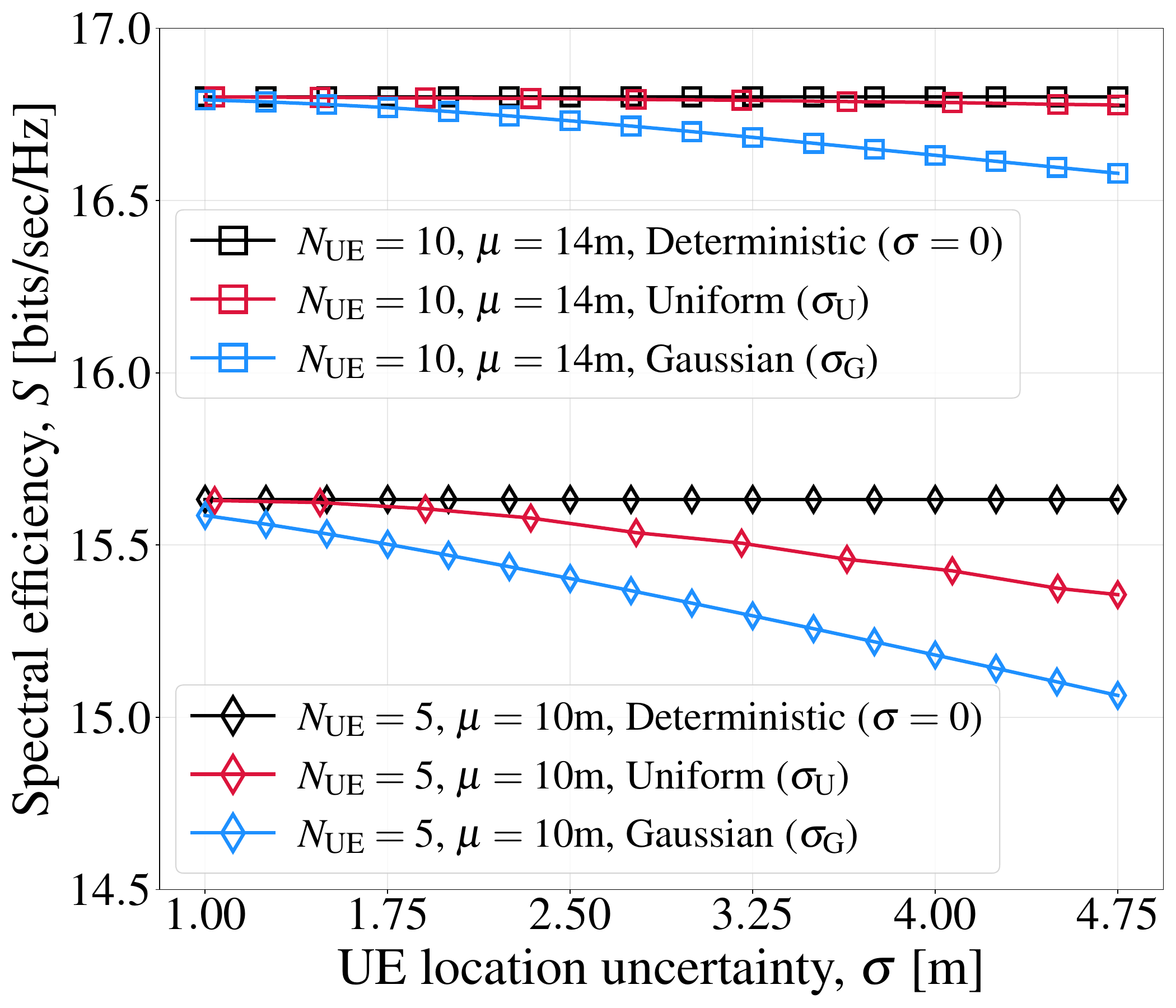}
\label{fig:SEvssig}}
\caption{SE trends under deterministic, Gaussian, and uniform UE-location scenarios. 
(a) SE versus frequency for \(N_{\rm AP}\in\{300,500\}\) and 
\(N_{\rm UE}\in\{1,20\}\) at \(\mu=10\) m. 
(b) SE versus mean UE location for fixed 
\(\sigma_{\rm G},\sigma_{\rm U}\in\{1,4\}\) m. 
(c) SE versus location uncertainty for fixed $\mu \in \{10,14\}$~m. 
For (b)--(c), \(N_{\rm AP}=500\), \(N_{\rm UE}\in\{5,10\}\), and 
\(f=300\) GHz.}
\label{fig:perfeval}
\vspace{-14pt}
\end{figure*}

\section{Numerical Analysis}\label{sec:Perf_ana}
This section evaluates the proposed closed-form Bessel beam configurations through numerical simulations. We first establish agreement with exhaustive search by (i) comparing the closed-form prediction of the optimal cone angle $\theta^*$ to the exhaustive-search benchmark and (ii) quantifying the resulting accuracy via empirical CDFs of the \ac{se} error over a broad set of system parameters. With this validation in place, we then use the closed-form rules to study \ac{se} trends and robustness versus frequency, \ac{ue} mean location, and location uncertainty.

\subsection{Closed-form approximation vs exhaustive search}
We begin by examining the behavior of the optimal ``cone angle'' $\theta^*$. Fig.~\ref{fig:theta_vs_nap} compares $\theta^*$ obtained via exhaustive search~\eqref{eq:exh_search} with the values predicted by the closed-form expressions~\eqref{eq:perfectue}, \eqref{eq:gauss}, and \eqref{eq:uni} for the three \ac{ue} location scenarios. We fix $f=300$~GHz and $N_{\rm UE}=1$, and vary the \ac{ap} array size by sweeping $N_{\rm AP}\in[400,700]$ antenna elements. For the uncertain-location cases, we set $\sigma_{\rm G},\sigma_{\rm U}=2$~m and choose $\mu=0.5\!\left(d_{\rm RA}+d_{\rm FA}/10\right)$ to represent a nominal \ac{ue} distance and uncertainty level. Note that exhaustive search is performed over the same $\Theta$ and $z$ grids as in Section~\ref{sec:uncertaincform}.

Fig.~\ref{fig:theta_vs_nap} shows that the closed-form solutions closely track exhaustive search under both Gaussian and uniform uncertainty. In contrast, the deterministic-location reference consistently yields a smaller ``cone angle'', highlighting the importance of the proposed correction factors for uncertain \ac{ue} location.

Having verified that the closed-form expressions accurately capture the optimal configuration variable $\theta^*$, in Fig.~\ref{fig:CDF}, we next evaluate its impact on the achievable \ac{se} performance. We fix the \ac{ap} size to $N_{\rm AP}=500$ and consider frequencies $f\in\{100,300\}$~GHz to demonstrate that the proposed framework applies to (sub-)THz regimes. 
The \ac{ue} mean location is selected within the radiative near-field region for all three \ac{ue} models,
while the uncertainty ranges are swept as $\sigma_{\rm G},\sigma_{\rm U}\in[1,5]$~m for the Gaussian and uniform error scenarios.

The accuracy of the closed-form approximation is quantified through the percentage SE error $\Delta S^{\%}_{\rm Dist}$, for $\mathrm{Dist}\in\{\rm D,G,U\}$, representing deterministic location, Gaussian error, and uniform error in location:
\begin{equation}
\Delta S^{\%}_{\rm Dist}(\mu,\sigma)\triangleq 100 \times 
\frac{S^{\rm Exh}_{\rm Dist}(\mu,\sigma)-S^{\rm CF}_{\rm Dist}(\mu,\sigma)}
{S^{\rm Exh}_{\rm Dist}(\mu,\sigma)}.
\end{equation}
Here, $S^{\rm Exh}_{\rm Dist}(\cdot)$ and $S^{\rm CF}_{\rm Dist}(\cdot)$ denote the exhaustive search and closed-form spectral efficiencies, respectively. The resulting CDF curves provide a rigorous statistical comparison between the proposed approximations and the optimal benchmark across varying uncertainty levels and system parameters.

We assess scalability with varying \ac{ue} array size by computing $\Delta S^{\%}_{\rm Dist}$ for $N_{\rm UE}\in\{1,5,10\}$. Fig.~\ref{fig:CDF}, shows the empirical CDFs of $|\Delta S^{\%}_{\rm Dist}|$ on a log-scaled $x$-axis; crosses mark the $95$th-percentile errors (the $0.95$-CDF intercepts) for each $(N_{\rm UE},f)$. In all three UE location scenarios, the CDFs are clustered with no long tails, indicating consistently high accuracy without any rare large-error events. For deterministic location (see Fig. \ref{fig:P_CDF}), the CDFs also exhibit the mild frequency dependence predicted in Section~\ref{sec:axial_peak}: errors are slightly smaller at $300$~GHz than at $100$~GHz, yet both are extremely small. The maximum $95$th-percentile errors across all tested configurations are only $0.057\%$ (deterministic), $0.00722\%$ (Gaussian), and $0.00471\%$ (uniform), confirming close tracking of the exhaustive-search optimum. Hence, validating that our approximations accurately capture the dominant first-order variations that affect the \ac{se} performance for the three \ac{ue} location scenarios.

\subsection{Performance dynamics of THz links using Bessel beams}

Fig.~\ref{fig:perfeval} examines the absolute SE performance and robustness of the proposed closed-form approximations. Throughout this subsection, we use \(P_{\rm tx}=0\) dBm and compute the noise power as \(n_{\rm F}=-174+10\log_{10}(B)+NF=-82\) dBm, with \(B=100\) MHz and noise figure \(NF=12\) dB, representative of THz receivers. We first consider the deterministic \ac{ue} case in Fig.~\ref{fig:SE}, where operating frequency is swept from $300$~GHz to $675$~GHz, and the \ac{se} is evaluated at mean \ac{ue} location $\mu=10$~m. We consider $N_{\rm UE}\in\{1,20\}$ and $N_{\rm AP}\in [300,500]$.

Fig.~\ref{fig:SE} shows the resulting \ac{se} versus $f$ plot for deterministic \ac{ue} location. We observe that increasing $N_{\rm UE}$ or $N_{\rm AP}$ provides a consistent \ac{se} gain across all $f$, showing that increasing aperture delivers the dominant array gain in this regime. As the frequency increases, we see a gradual decrease in achieved \ac{se}. This is because, with fixed antenna element spacing and $N_{\rm AP}$, increasing $f$ reduces the physical aperture and $Z_{\max}$, reducing the array gain and tightening the axial region, resulting in the gradual \ac{se} decrease.

Next, we analyze the impact of uncertain \ac{ue} location in Fig.~\ref{fig:SEvsU}--\ref{fig:SEvssig} on achieved \ac{se}. We sweep $\sigma_{\rm G},\sigma_{\rm U}\in[1,5]$~m and $\mu \in[9.25,14.75]$~m fixing $f=300$ GHz, $N_{\rm AP}=500$, and report results for $N_{\rm UE}\in\{5,10\}$. Fig.~\ref{fig:SEvsU} plots \ac{se} versus $\mu$ for fixed $\sigma \in \{1,4\}$~m (deterministic location reference), and Fig.~\ref{fig:SEvssig} plots \ac{se} versus $\sigma$ for fixed $\mu \in \{10,14\}$~m.

In Fig.~\ref{fig:SEvsU}, increasing $\mu$ yields a gentle \ac{se} decay consistent with the extended $Z_{\max}$ depth of Bessel beams; the deterministic-uncertain gap shrinks slightly at larger distances, and larger $N_{\rm UE}$ reduces the uncertainty penalty via an averaging effect observed due to coherent combining. While in Fig.~\ref{fig:SEvssig}, all three models nearly coincide at small uncertainty, with the Gaussian and uniform curves collapsing to the deterministic location baseline, i.e., smooth convergence to \eqref{eq:perfectue}. As uncertainty grows, the stochastic cases diverge gradually but remain smooth and monotonic. 

Overall, Fig.~\ref{fig:perfeval} highlights that the \ac{se} increases consistently with larger $N_{\rm AP}$ and $N_{\rm UE}$ and decreases gradually with $f$ for the deterministic case. Under uncertain \ac{ue} location \ac{se} decays gently as $\mu$ increases, with the uncertainty penalty shrinking at larger distances. Meanwhile, increasing the standard deviation $\sigma$ leads to a performance degradation and a growing divergence from the deterministic \ac{ue} reference.

\section{Conclusion}{\label{sec:Con_fut}}

This paper studied phase-only Bessel beam configuration for radiative near-field (sub-)\ac{thz} downlink \ac{mimo} links under deterministic and uncertain \ac{ue} location knowledge. We formulated the Bessel phase configuration as an expected \ac{se} maximization problem and approximated the solution using the properties of the axial power profile of Bessel beams. Our analytical contribution is a set of $\mathcal{O}(1)$-complexity closed-form expressions that map the near-optimal Bessel beam configuration directly to available \ac{ue} location information. 

Our numerical study confirmed that the derived configurations closely track exhaustive search across a range of array sizes, frequencies, and \ac{ue} location scenarios. The evaluations also highlighted that the deterministic--uncertain \ac{se} gap shrinks at larger distances but gets larger as uncertainty in the \ac{ue} location grows. These closed-form expressions and the presented analysis facilitate the design of future practical near-field \ac{thz} systems (e.g., codebook design), as well as accurate yet analytically tractable system-level studies of near-field \ac{thz} networks, e.g., capacity, coverage, and interference modeling.

\section*{Acknowledgment}
This work has been supported by SSF grants ID24-0074, FFL-9 (V. Petrov), and FUS21-0004 (SAICOM; G. Fodor).
\balance
\bibliographystyle{IEEEtran}
\bibliography{ref}

\end{document}